\documentclass[lettersize,journal]{IEEEtran}
\usepackage{amsmath,amsfonts}
\usepackage{algorithmic}
\usepackage{algorithm}
\usepackage{array}
\usepackage[caption=false,font=normalsize,labelfont=sf,textfont=sf]{subfig}
\usepackage{textcomp}
\usepackage{stfloats}
\usepackage{url}
\usepackage{verbatim}
\usepackage{graphicx}
\usepackage{cite}
\usepackage{colortbl}
\usepackage[table]{xcolor}
\usepackage{makecell}

\begin{document}
	\title{{Single-Photon Counting Receivers for Optical Wireless Communications in Future 6G Networks}}
	\author{Shenjie Huang, Danial Chitnis, Cheng Chen, Harald Haas,~\IEEEmembership{Fellow,~IEEE}, Mohammad-Ali Khalighi,~\IEEEmembership{Senior Member,~IEEE}, Robert K. Henderson,~\IEEEmembership{Fellow,~IEEE}, and Majid Safari,~\IEEEmembership{Senior Member,~IEEE}
	\thanks{S. Huang, D. Chitnis, Robert K. Henderson, and M. Safari are with the School of Engineering, the University of Edinburgh, Edinburgh EH9 3JL, UK (e-mail: shenjie.huang@ed.ac.uk). C. Chen and H. Haas are with LiFi Research and Development Centre, University of Strathclyde, Glasgow G1 1RD, UK. M. A. Khalighi is with Aix-Marseille University, CNRS, Centrale Marseille, Institut Fresnel, Marseille, France.}
	\thanks{This work was supported by Engineering and Physical Sciences Research Council (EPSRC) under Grant EP/S016570/1 (TOWS). For the purpose of open access, the author has applied a Creative Commons Attribution (CC BY) licence to any Author Accepted Manuscript version arising from this submission.}}	

	\maketitle
	
	\begin{abstract}
	Optical wireless communication (OWC) offers several complementary advantages to radio-frequency wireless networks such as its massive available spectrum; hence, it is widely anticipated that OWC will assume a pivotal role in the forthcoming sixth generation wireless communication networks. Although significant
	progress has been achieved in OWC over the past decades,
	the outage induced by occasionally low received optical power
	continues to pose a key limiting factor for its deployment. In this work, we discuss the potential role of single-photon counting (SPC) receivers as a promising solution to overcome this limitation. We present an overview of the applications of SPC-based OWC systems in 6G networks, {introduce their major performance-limiting factors, propose a performance-enhancement framework to tackle these issues,} and identify critical areas of open problems for future research.
	\end{abstract}
	
	
	\section{Introduction}
    The rapid emergence and evolution of smart devices, interactive services, and intelligent applications have generated a substantial demand for high-speed wireless communication.  
    Researchers have initiated investigations to explore the potential characteristics and capabilities of the future sixth generation (6G) networks. The forthcoming 6G technology is anticipated to offer essential features encompassing ultra-high speed, sub-centimetre geo-location accuracy, global coverage, ultra-high density, enhanced energy efficiency, ultra-low latency, and high intelligence levels \cite{You6G}. These key features will greatly accelerate the development of newly emerged applications such as metaverse, autonomous vehicles, smart city, and extended reality. It is predicted that all available spectra, including sub-6 GHz, millimeter wave, terahertz, and optical frequency bands, will be exploited in 6G. {Therefore, optical wireless communication (OWC) is envisioned to play a crucial role in future wireless networks owing to its numerous advantages, including ultra-wide bandwidth, inherent security, simple beamforming implementation, low energy consumption and cost, insensitivity to radio frequency (RF) electromagnetic interference, and potential to achieve high-resolution sensing.}
	
    OWC was primarily utilized in military applications until a few decades ago, when substantial research endeavours were undertaken, aiming to expedite its commercialization. The primary technologies that have undergone extensive investigation include terrestrial and space free-space optical communication (FSO), indoor visible light communication (VLC), optical camera communication, and underwater OWC. {Despite having three orders of magnitude more spectrum resources compared to its RF counterpart, OWC encounters notable challenges that need to be addressed prior to its widespread deployment.} In particular, the occasional outages arising from fluctuations in the received optical power constitute a prominent concern. Such power fluctuations can be attributed to user mobility,  beam blockage, device orientations, and dimming in indoor VLC, and to atmospheric turbulence, adverse weather conditions, and transceiver vibrations in outdoor FSO. 
	
\begin{figure}[!t]	\centering\includegraphics[width=0.49\textwidth]{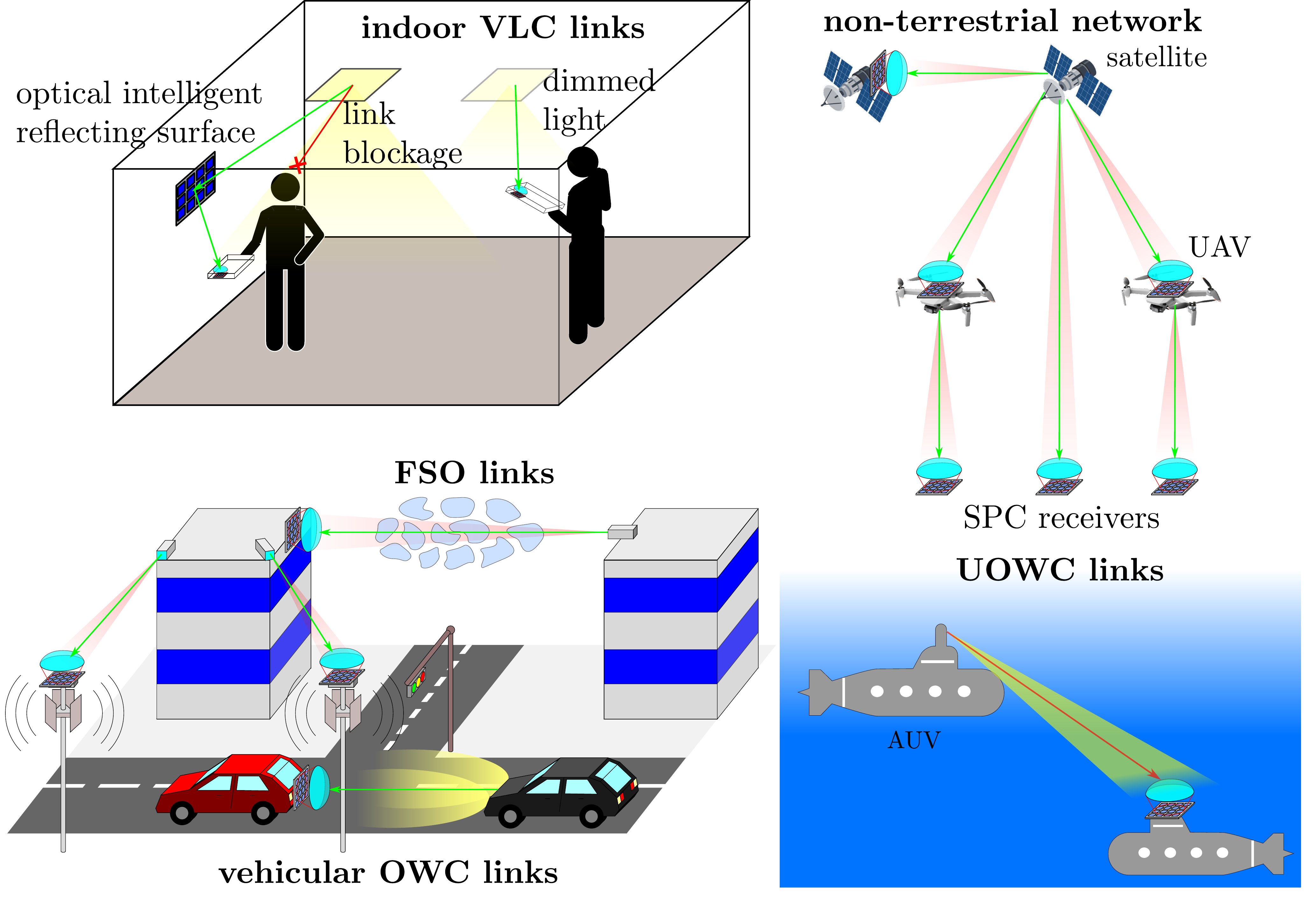}   
\caption{The applications of SPC-based OWC in 6G.}\label{SPC6G}
\end{figure}

    A promising solution to improve the OWC link reliability relies on the utilization of single-photon counting (SPC) receivers. The gains of typical SPC receivers are normally on the order of millions, while the gains of traditional positive-intrinsic-negative (PIN) photodiode (PD) and avalanche photodiode (APD) are only unity and on the order of hundreds, respectively \cite{Zhang18}.  Therefore, the achievable sensitivity of SPC receivers is significantly higher than the widely employed conventional photodetectors and could approach the quantum limit, which represents the fundamental minimum optical power required to fulfil specific communication requirements \cite{Zimmermann19}.  
    The SPC technologies which have been widely investigated in the literature include photomultiplier tube (PMT), single-photon avalanche diode (SPAD), and superconducting nanowire single-photon detector (SNSPD). SPC detectors have been used in a variety of other applications where the detection of extremely weak optical signals is demanded such as light detection and ranging, quantum cryptography, and time-of-flight imaging. In recent years, the utilization of this technology in OWC has garnered substantial attention from both academia and industry. Some possible 6G applications of SPC-based OWC are illustrated in Fig. \ref{SPC6G}. {It is essential to highlight that, in practical implementation, the susceptibility of SPC receivers to background light can be effectively mitigated through the adoption of various approaches, such as employing optical filters with narrow passbands and reducing the receiver field-of-views (FOVs). }

	{This article aims to deliver a comprehensive overview of SPC-based OWC and explore its potential in 6G networks. To this end, we first introduce common technologies for the implementation of SPC receivers, with a particular emphasis on SPAD arrays, which recently attracted significant interest. Following that, the recent advances in SPC-based OWC systems are presented. We report a new achievable rate of more than $10$ Gbps for SPC-based OWC, which further demonstrates its potential for applications with medium-to-high data rate requirements. Next, the performance impairments of the SPC-based OWC system are introduced. To enhance the performance and thus expand the potential of SPC receivers in 6G networks, a general framework is proposed to effectively mitigate the impact of the receiver nonidealities and inherent noise signal-dependency. Finally, we present some potential directions for future research. }

	\section{{Single-Photon Counting Technologies}}
Most current OWC systems employ PIN PDs or APDs at the receiver.
APD requires a higher reverse bias close to its breakdown voltage and can provide an internal gain ranging from $10$ to $200$.
In contrast, typical SPC receivers can provide considerably higher internal gains, resulting in significantly improved sensitivities. The ability to manufacture large arrays of SPC detectors such as SPAD and SNSPD is another important factor that has crucially enhanced their efficiency to practical levels. Table \ref{table} compares the main features of PIN PDs, APDs, and SPC detector technologies, which will be discussed next. 
This table also clearly demonstrates the substantial superiority of SPC arrays over individual detectors in terms of active area, signal dynamic range, and ambient light immunity.  

	\newcolumntype{s}{>{\columncolor{gray!30}} m{1.6cm}}
	\begin{table}
		\renewcommand{\arraystretch}{2.2}
		\caption{Comparison of main features of considered detectors}
		\label{table}
		\centering
		\resizebox{0.48\textwidth}{!}
		{\begin{tabular}{|s||c|c|c|c|c|}
				\hline
				\rowcolor{gray!30} Parameter & PIN PD & APD & PMT & SPAD &  SNSPD \\
				\hline\hline
				 Sensitivity & low & moderate & very high & very high & very high\\
				\hline
				 Gain & 1 & $10$-$200$ & $10^6$-$10^7$ & $10^6$ & no multiplication\\
				\hline
				{Active area} & moderate & moderate & very large & \makecell{small\\ (large with array)} & \makecell{small\\ (large with array)}\\
				\hline
				{Detection $\,\,$ efficiency} & high & moderate & low & moderate & high \\
				\hline
				Dynamic range & very large & very large & moderate & \makecell{small\\ (large with array)} & \makecell{small\\ (large with array)} \\
                    \hline      
            Ambient light immunity & high  & moderate & low & \makecell{low\\(moderate with array)}  & \makecell{low\\(moderate with array)} \\
				\hline
				Photon $\,\,\,\,\,$ timing jitter & - &-  & \makecell{moderate\\ $50$ ps-$1000$ ps} & \makecell{low\\ $20$ ps-$300$ ps}  & \makecell{very low \\ $3$ ps-$60$ ps}\\
				\hline
				Temperature sensitivity & low & moderate  & low  & moderate  & {very high}
				 \\
				\hline
				{Cost}& {low} & {moderate}  & {high}  & {low}  & {very high}
				\\
				\hline
				SWaP \hspace{5mm}(size, weight and power) & low & moderate & very high & moderate & \makecell{very high \\(incl. cooling system)} \\
				\hline
		\end{tabular}}		
	\end{table}

	\subsection{PMT and SNSPD}
	PMTs are the most conventional SPC detectors and have been employed since the era preceding semiconductors. PMTs consist of two primary electrodes within the vacuum tube, with a high electrical field typically on the order of kilovolts applied between the anode and cathode. 
 PMTs have the largest photodetection area among photodetectors, ranging from $1$ cm to $10$ cm in diameter, which enables a large receiver FOV and high power gain. They are also advantageous in terms of high gain (typically $10^6-10^7$) and low noise level. However, PMTs have some practical disadvantages, including their bulky size, and high operating voltage and cost, which limits the suitability of PMTs for some 6G applications. {Nonetheless, they retain their commercial value in some fixed point-to-point OWC applications, such as underwater applications. This is primarily due to their notably large active area, which significantly augments sensitivity and alleviates the alignment challenges.}
	
	In contrast to PMT, SNSPD is a newly emerged SPC detector, which has advanced rapidly over the last decade.  The detector element is a thin and narrow superconducting nanowire. The nanowire should operate at low temperature range of $1.5-4$ K, well below the superconducting transition temperature, and is biased just below its critical current. When a photon arrives at the nanowire, an instability of the superconducting state emerges which leads to a local resistive hotspot. The increasing resistance, which is on the order of several k$\Omega$, redirects the current and triggers a fast voltage pulse that can be amplified and measured.  
	The key advantages of SNSPD include high detection efficiency (above $90 \%$) and short recovery time (potentially less than $1$ ns)
	\cite{Korzh20}. However, the extremely low operation temperature requires the development of more advanced cost-effective cooling technologies. 
	\subsection{SPAD}
	SPAD detectors, which can be implemented as large arrays using {complementary metal–oxide–semiconductor (CMOS)} processes, have recently attracted significant interest. Compared to other SPC receivers, SPADs also offer other advantages including small structural dimensions, reduced weight, and lower operation voltage.
	These inherent attributes make SPAD-based OWC systems particularly attractive for 6G mobile wireless networks.
	
	SPADs can generate a digital pulse when detecting a single photon eliminating the need for complex analog circuitry for signal amplification. Employing a similar structure as APD, SPAD is biased above the breakdown voltage and operates in Geiger mode, ensuring that the avalanche process generates an extensive charge. The gain of SPAD is similar to PMT and is dramatically higher than APD. This large internal gain results in a large enough output voltage that can be interfaced with standard digital circuits, such as pulse counters. 
	
	The operation of a SPAD requires the use of a quenching circuit, which has the role of stopping the avalanche current generated by an incoming photon or dark noise, while a reset circuit prepares the avalanche diode for the next detection. The quenching and reset processes are typically achieved via passive or active circuits. Passive quenching (PQ) employs a resistor  to reduce the bias voltage across the diode as soon as the avalanche current is generated, thus quenching the avalanche process. 
	 In active quenching (AQ), the quenching and reset processes are accelerated through the use of active components for expediting the avalanche process and establishing a detector \textit{dead time} that allows charges to vacate the avalanche region and resets the junction capacitance to the initial voltage.

	Although both PQ and AQ SPADs exhibit excellent linearity at relatively low light intensities, they have distinct counting profiles at high intensities. AQ SPAD features a fixed dead time period when no incident photon can be detected. Therefore, with the increase of the light intensity, the detected photon count rate saturates. 
	PQ SPAD has the paralysis property leading to even a reduction of count rate at very high intensities. Note that, unlike AQ SPAD, PQ SPAD is not totally inactive during its recovery time, but the term dead time is still commonly used for the period it takes after a detection event until the detector recharges to the pre-avalanche bias voltage. Although AQ SPADs present higher resilience to intensities at higher levels and typically have lower dead time, they require more complex timing circuitry, resulting in higher cost and larger size.  {Consequently, simpler PQ SPADs have predominantly found utilization in recent OWC experiments. } {AQ SPADs, on the other hand, are more suitable for those applications where substantial data throughput is needed. Directing further research efforts toward cost reduction for AQ SPADs can substantially expand their range of potential applications.} 
	
	In addition to the dead time effect which is a major aspect of SPAD receivers when operating at high-speed systems, the other characteristics of SPAD array receivers need to be considered include dark count rate (DCR), photon detection efficiency (PDE), afterpulsing, and crosstalk \cite{He21,Zimmermann17}.

    \textbf{SPAD Arrays:} Single SPADs have photosensitive diameters typically ranging from $10$ to $100$ $\mu$m. 
    Therefore, SPAD array detectors, 
    also commercially known as multi-pixel photon counters or silicon photo-multipliers,
    are typically adopted to attain greater detection areas and wider linear dynamic ranges. There are two primary methods to read the output of a SPAD array: summing the avalanche currents generated by each SPAD cell or digitally combining each SPAD output using logic gates. 
 The first method has less complexity due to analog summation and is widely adopted in commercial detectors. However, the readout circuitry requires amplification to account for the small output voltage, which introduces analog noise. 
 In contrast, the second method provides a clear digital output, and the pulses are counted similarly to a single SPAD. Nevertheless, the implementation complexity significantly increases to maintain the digital summation of all photons through logic gates. As a result, despite their superior performance, SPAD arrays with such a combining method are predominantly limited to research and development (R\&D) applications. {Therefore, in future applications, it is crucial to consider the trade-off between performance and complexity when selecting the combining techniques. }   
	
    \section{{Recent Advances in SPC-Based OWC}}
As discussed in the previous section, SPAD is the most promising SPC technology for integration into OWC systems in the upcoming 6G era. 
Over the last decade, there has been a substantial body of experimental research  focused on exploring the potential applications of SPAD in various OWC systems. In particular, sensitivities of $-55.7$ dBm and $-51.6$ dBm at $50$ Mbps and $100$ Mbps, respectively, were achieved by a $0.35$ $\mu$m SPAD CMOS array receiver with an array size of $4$ and a dead time of $9$ ns \cite{Zimmermann17}. With the same array size, another SPAD receiver with a dead time of only $3.5$ ns was implemented which reached a sensitivity of $-43.8$ dBm at $200$ Mbps \cite{Steindl18}. In addition, a $60$-SPAD array receiver with a minimum dead time of $5$ ns was reported in \cite{Zhang18} which achieved $-53$ dBm sensitivity at $100$ Mbps. 

Besides the aforementioned results achieved by R\&D SPAD receivers,  a considerable body of literature has also examined the use of commercial off-the-shelf SPAD receivers in OWC. 
These receivers typically feature larger array sizes and higher PDE
at the expense of increased noise levels. The gigabit SPAD-based OWC was first reported in \cite{Ahmed20}.  
Later, an achievable data rate of $3.45$ Gbps was attained by using a commercial receiver with an array of $14,410$ SPADs
\cite{Matthews21}.  Using the same receiver, a data rate of $5$ Gbps was achieved through the utilization of optical orthogonal frequency division multiplexing (OFDM) \cite{Huang:22}. More recently, $8$ Gbps data transmission of SPAD-based OWC with wide-band vertical cavity surface emitting laser array was reported in \cite{Liu23}. 

\begin{figure*}[!t]	\centering\includegraphics[width=0.89\textwidth]{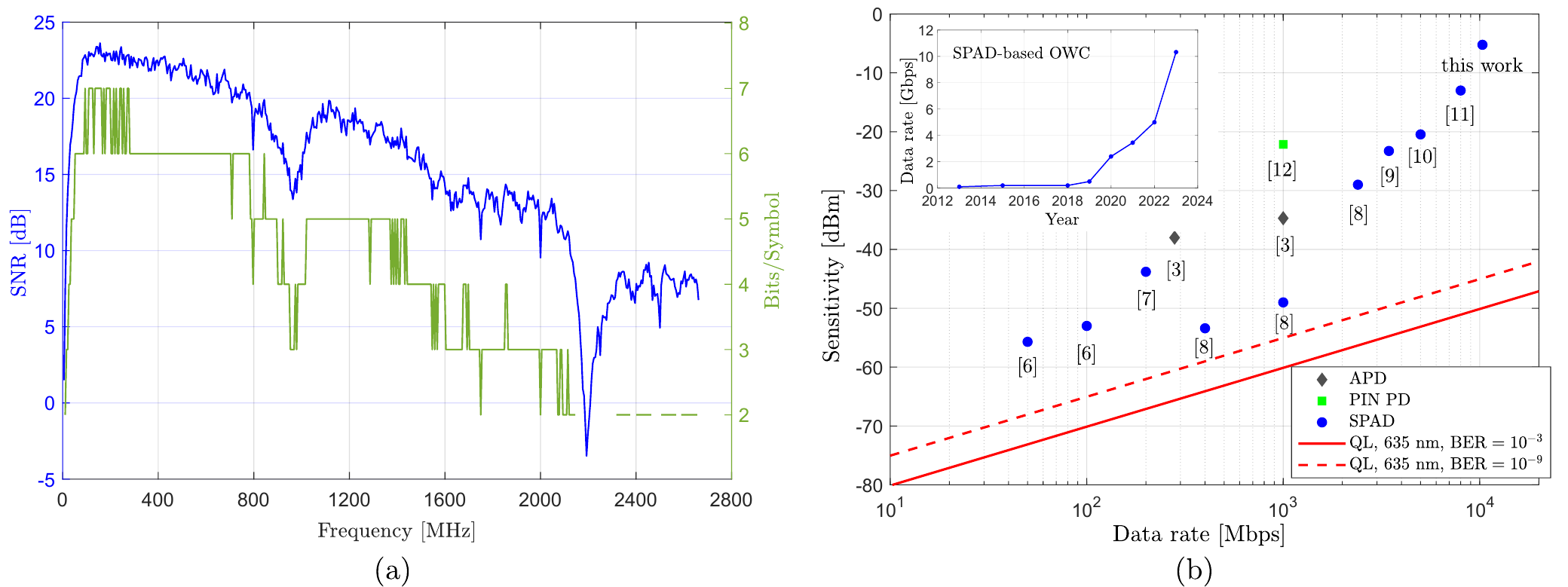}\vspace{-3mm}
	\caption{{Achievable rate of SPAD-based OWC: a) measured SNR and adaptively loaded bits on each OFDM subcarrier to achieve $10.33$ Gbps; b) reported sensitivity versus the data rate achieved in the current literature including that of the system implemented in this work, QL: quantum limits.} }\label{quantum_limit}
\end{figure*}

{To further improve the achievable data rates of SPAD-based receivers and assess their potential to meet the 6G key performance indicators, we employ a SPAD-based OWC setup similar to our previous work in \cite{Huang:22}, while utilizing a different off-the-shelf SPAD receiver, i.e., Onsemi C-Series 10010, which possesses a small dead time and a wide bandwidth attributed to its small $10$ $\mu$m pixel size. By employing the optical OFDM modulation together with adaptive bit and energy loading, and mitigating the dead time effects through nonlinear equalization and waveform peak-to-average power ratio optimization, a data rate of $10.33$ Gbps at a bit error rate (BER) less than $3.8\times 10^{-3}$ is achieved under a received optical power of $-5.2$ dBm. The measured signal-to-noise ratio (SNR)
and bit loading results are presented in Fig. \ref{quantum_limit}a.
To the best of the authors' knowledge, this is the first time a SPAD-based OWC system with data rates beyond $10$ Gbps is reported.}

Figure \ref{quantum_limit}b shows the sensitivity of SPAD-based receivers reported in the literature versus the achievable data rate along with that for some benchmark PIN PD \cite{Ju16PIN} and APD \cite{Zimmermann19} receivers. In low data rate regimes, the superiority of SPAD-based OWC over APD and PIN PD in terms of sensitivity is clearly demonstrated, which is close to the quantum limit \cite{Zimmermann17}. In contrast, in high data rate regimes, their sensitivity levels degrade primarily due to the inherent nonideality of SPADs and the limited array sizes for efficiently designed R\&D SPADs. However, it is anticipated that further advancements in SPAD OWC, including SPAD device manufacturing and novel dead time effect mitigation techniques, will soon bring a restoration of its superiority in higher data rate regimes. This figure also illustrates the remarkable increase in the measured data rate of SPAD-based OWC systems over recent years. According to this trend, a further improvement in achievable data rates is highly plausible in the near future.          

	\section{{Efficient Design of SPC-Based OWC Systems}}
	
	\subsection{SPC Impairments}\label{DeadTimeeffects}
	Among the various nonidealities associated with SPC receivers, the dead time causes nonlinear effects  that are particularly challenging to model, while  other nonidealities can be adequately described using basic models such as a linear power loss (e.g., PDE, or fill factor) or an increase in additive noise level (e.g., DCR, or afterpulsing), resulting in a reduction of SNR. In fact, the dead time effect induces a fundamental limit on the achievable data rate of OWC systems. It is worth noting that the impact of dead time on OWC performance is not exclusive to SPAD receivers but also applies to any SPC-based receivers with a recovery time significant compared to the communication bandwidth. The dead time effect hampers the SPC receiver's ability to respond to all incident photons thereby reducing the detected photon count, which degrades communication performance in two ways: 1) \textit{nonlinear response} to incident photons  (e.g., saturation or gain reduction), and 2) \textit{dead-time-induced inter-symbol-interference} (ISI). These two nonlinear effects are both caused by dead time and are thus correlated, but we will discuss them separately next.

	\begin{figure}[!t]	\centering\includegraphics[width=0.48\textwidth]{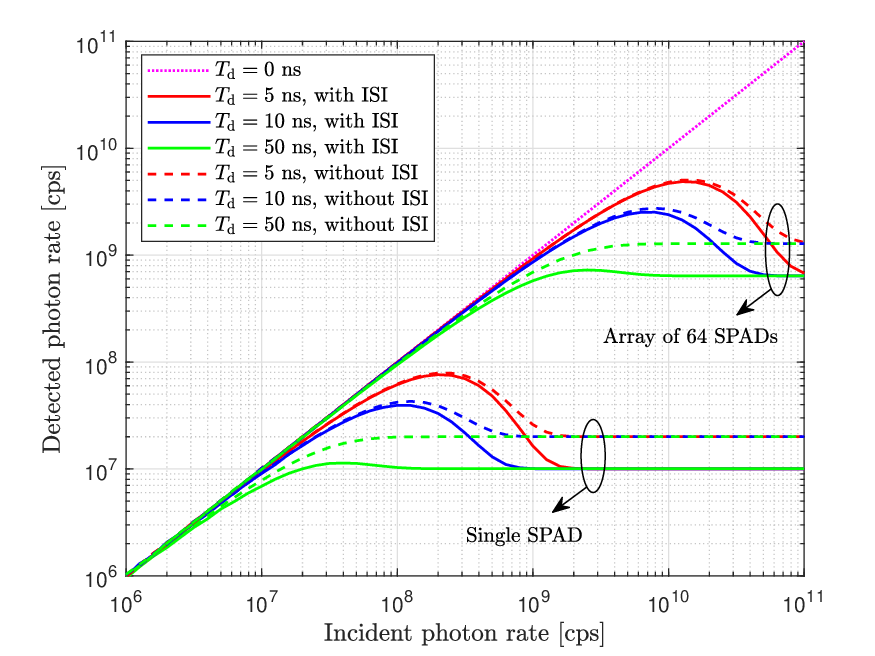}
	\caption{The detected photon rate versus the incident photon rate for the SPAD receivers with various dead time $T_\mathrm{d}$ in the presence and absence of dead-time-induced ISI. On-Off keying (OOK) modulation scheme is employed and the SPAD is PQ based. The presented photon rates refer to bit `1' reception. A symbol time of $50$ ns is considered and a negligible incident photon rate for bit `0' reception is assumed.}\label{photon_rate}
\end{figure}

\begin{figure*}[!t]		\centering\includegraphics[width=1.01\textwidth]{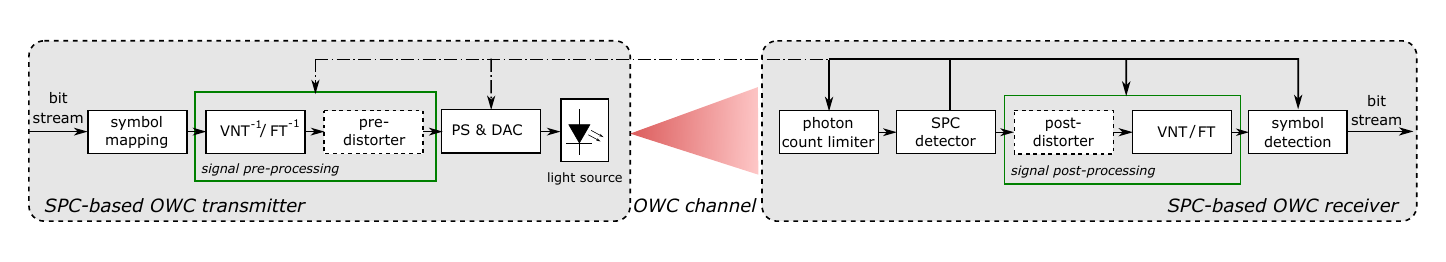}\vspace{-3mm}
	\caption{{Schematic diagram of the proposed SPC-based OWC framework, VNT: variance normalizing transform; FT: Fourier transform; PS: pulse shaping; DAC: digital-to-analog converter.}}\label{new_system}
\end{figure*}	 

\textbf{Nonlinear Response:}  
Due to the existence of dead time, any photons arriving at an SPC detector within the dead time period of a successful detection of the same detector would be missed. {As the incident photon rate increases, the frequency of photon arrivals also increases, leading to shorter photon inter-arrival times, thereby exacerbating the loss of photon detection.}  
Fig. \ref{photon_rate} presents an example of the nonlinear response of a PQ-based SPAD receiver, showing the detected photon rate versus the incident photon rate.
{This result is attained by following steps.  A random data stream is generated based on which a photon arrival sequence received by each SPAD can be generated according to the theory of Poisson process. Next, the photon arrival sequences are filtered throughout by the dead time. The photon counts detected by the array for different symbols are then recorded, and subsequently, the average detected photon rate is calculated. The above procedure can exactly mimic the photon reception in practical SPAD receivers and has been extensively adopted in numerous works \cite{Huang20JLT}.}
We observe that, for a receiver with a single SPAD, the detected photon rate initially increases linearly as the incident photon rate increases, but then begins to decrease before eventually reaching a saturation point. This differs from the ideal SPC receiver without dead time, whose detected photon rate increases linearly with the incident rate. In addition, we can observe that an increased dead time exacerbates the loss in photon counts it induces. Furthermore, Fig. \ref{photon_rate} 
presents the effectiveness of employing larger SPAD arrays in combating the dead time nonlinear response.

	\textbf{Dead-Time-Induced ISI:} The dead time started in a symbol may extend to the subsequent symbols, introducing photon counting blocking in one or multiple subsequent symbols depending on the ratio between dead time and symbol duration. In effect, SPC detector might be inactive at the beginning of the photon counting period of the symbol due to the last photon detection in previous symbols, resulting in a nonlinear dead-time-induced ISI effect. This ISI effect is inherently different from the ISI effects that impair traditional bandwidth-limited communication systems.    
	ISI in conventional communication systems can be precisely characterized as the linear or nonlinear combination of a set of neighbouring transmitted symbols, making it able to be effectively equalised. However, this does not hold true for ISI induced by dead time. Dead-time-induced ISI directly depends on the arrival time of the last photon detection before the start of the considered symbol and is only statistically related to the transmitted symbols \cite{Huang20JLT,He21}. Since the instantaneous block period is inherently random, the considered ISI exhibits an additional source of noise, which is more severe when the transmission rate and the received optical power are relatively high.  
	{To illustrate the impact of ISI, Fig. \ref{photon_rate} also depicts the photon count rates when dead-time induced ISI is ignored, which are achieved based on the assumption that SPADs are ready to collect photons at the beginning of each symbol. It illustrates the regimes where ISI effects are significant, requiring the adoption of ISI mitigation techniques.}

\subsection{{Proposed Performance Enhancement Framework}}\label{PerfEnhance}
{To improve the reliability and achievable data rate of SPC-based OWC systems at different power regimes, we propose a general framework 
as illustrated in Fig. \ref{new_system}. This framework incorporates signal pre-processing and post-processing blocks at the transmitter and receiver, respectively, which are designed to effectively mitigate the impact of receiver dead time and noise signal-dependency. Its key features are introduced as follows.} 

{\textbf{Noise Normalization:}
In contrast to conventional PDs, the noise characteristics of SPC receivers are notably more complex, which is intricately related to both signal strength and dead time. Note that there are specific transformations that can be applied to the output of a signal-dependent noise channel to effectively decouple signal from noise and approximately convert it to a conventional signal-independent noise channel. To achieve this,  the encoding and decoding of the signal needs to be performed in the transformed domain. In the case of single-carrier modulation, e.g., pulse-amplitude modulation (PAM), variance normalizing transform (VNT) techniques can be designed to turn the noise samples after transformation to be approximately independent of the signal \cite{huang2022spad}.  For multi-carrier modulation systems e.g., optical OFDM, the signal is inherently encoded/decoded in the frequency domain, which is defined by the Fourier transform (FT). Interestingly, FT can also provide an averaging effect that diminishes the signal dependency of the noise after the transform at receiver. Therefore, to implement noise normalization, the inverse of the corresponding transformation (VNT or FT) is employed at the transmitter and the direct transformation is used at the receiver side. }

{\textbf{Pre/Post-Distortion:} In the proposed system, the nonlinear effects caused by dead time, i.e., component nonlinear response and ISI, can be mitigated by pre-distorter at the transmitter or post-distorter at the receiver. While the pre-distortion approach may offer superior performance, it necessitates the acquisition of channel state information (CSI) at the transmitter in order to achieve decent performance, thereby introducing greater complexity into the system. The use of a combination of both distorters can also be considered if necessary. In addition to SPC receivers, other components within the system, such as the light source, may introduce either memory or memoryless nonlinear distortions, which also require pre- or post-distortion for compensation  \cite{Ying15}. In the proposed system, if nonlinearities stemming from other components prove to be significant, their mitigation can be seamlessly integrated into the distorter design. }

{\textbf{Photon Count Limiter:} To avoid excessive receiver nonlinear effects, a photon count limiter is also proposed at the receiver to limit the incident photon count to the SPC detector. This limiter may be implemented through two approaches: 1) Optically by employing a variable optical attenuator with adjustable transmittance via an input electrical signal, and 2) Electrically by using a gating circuit capable of raising or lowering the bias voltage of the detector, thus turning the detector ON and OFF in well-defined time intervals.}

{\textbf{CSI Utilization:} The SPC detector receives the optical signal and generates a digital photon count during each symbol, which is then used in subsequent processes. The optical power estimated by the detector can be utilized to control the photon count limiter and the following signal post-processing process. It may also be looped back to the transmitter to adaptively adjust the pulse shaping, thereby reducing the receiver dead time effects. Besides the photon count information, SPC detectors are able to provide precise photon arrival information thanks to their high temporal resolution. Such photon arrival information can be fed into the symbol detection process, facilitating the implementation of novel detection schemes that offer improved performance \cite{Huang20JLT}.}  
\begin{figure}[!t]		\centering\includegraphics[width=0.5\textwidth]{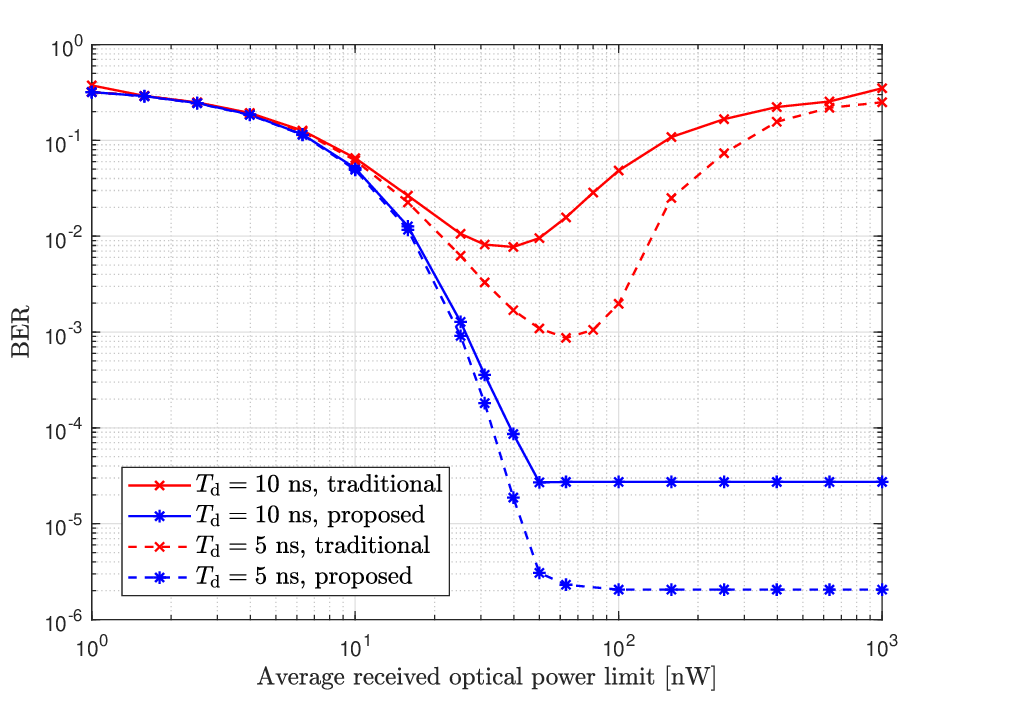}
	\caption{{Comparison of BER with the average received optical power limit for the proposed and the traditional system with different dead times. $4$-PAM modulation, SPAD array with $2048$ pixels, received background noise power of $10$ nW, symbol duration of $5$ ns, wavelength of $785$ nm, and PDE of $0.18$ are considered.}}\label{BER_proposed}
\end{figure}	

{To shed light on the performance improvement achieved by the proposed system, we conduct a numerical performance analysis of the proposed system with a SPAD array detector considering the traditional SPC system without pre- and post-signal processing and photon count limiter as a benchmark.  We utilize VNT and pre-distorter designed with the assumption of negligible dead-time-induced ISI \cite{huang2022spad}, alongside a Volterra series based equalizer as a post-distorter. Note that the assumption of negligible ISI is not accurate for high-speed systems operating at high received power levels, which makes the simulated system to be sub-optimal. Fig. \ref{BER_proposed} presents the BER performance of the considered SPC systems as a function of the received optical power limit. It can be observed that employing the proposed system, despite the use of sub-optimal blocks, results in a significant performance improvement. 
It is expected that an optimal system design (e.g., based on a data-driven approach) will yield further performance enhancement.}

\section{{Concluding Remarks and Future Research Directions}}
	{Although SPC-based OWC can achieve high sensitivity, its performance is limited by certain non-idealities, particularly the dead time effects. In this work, we have reported, for the first time, an experimentally demonstrated SPC-based OWC system with an achievable rate exceeding $10$ Gbps. Moreover, a general framework is proposed to effectively mitigate the dead time and the noise signal-dependency effects, leading to substantial BER improvement. 
    Nonetheless, great efforts are still required to overcome some challenges, thus facilitating the widespread adoption of SPC-based OWC in 6G networks. Below, we outline some prospective avenues for future research. }
	
	\subsection{Practical Models for SPC Receivers}
	The design and performance analysis of SPC-based OWC systems strongly depend on how well the SPC non-ideal effects, such as dead time, can be modelled. 
 For example, many existing studies in the literature have modelled the detected photon count of SPC receivers based on ideal Poisson distribution. 
 By disregarding the impact of dead time, this model significantly underestimates the {BER} performance. To this end, some research efforts have been devoted to developing analytical models of detected photon counts in the presence of dead time. However, to simplify mathematical derivations, the dead-time-induced ISI is normally ignored and the system is effectively treated as memoryless. {Fig. \ref{photon_rate} shows how ignoring ISI (dash curves) degrades the accuracy of the analysis}. Hence, the accurate modelling of the counting statistics of SPC receivers across a wide range of data rates and received power levels remains an open problem.

\subsection{Specification Improvements}
    The sensitivity gap between the practical SPC receivers and the ideal SPC receiver is mainly due to manufacturing imperfections.
    {For example, the practical SPAD array receivers suffer from limited fill factor, relatively high crosstalk probability and dead time, and count loss induced by the combining techniques.  Therefore, considerable efforts should be devoted to effectively improving the specifications of commercial receivers without incurring significant costs.}
    Developing SPC detectors with various materials and technologies would expand the range of highly sensitive receivers to new distinct bands. For example, the near-infrared (NIR) band is widely used in high-speed optical communication systems, making the design of highly efficient SPC receivers operating in this band of great importance. Unfortunately, current available SPC detectors with high PDE in the NIR band, such as InGaAs SPADs, are prone to high DCR and afterpulsing. 
    Therefore, further research is required to develop high-performance yet low-cost SPC receivers for NIR spectrum. {This may be achieved through the exploration of novel materials and manufacturing techniques such as the utilization of backside-illuminated process. It is worth noting that, recently, there has been a notable inclination towards the manufacturing of 3D-stacked SPADs,  which feature significantly enhanced PDE in NIR while maintaining a low DCR. Exploring the potential applications of these detectors in OWC domain represents promising research perspectives.}

    \subsection{{Signal Processing Techniques}}
    {The utilization of advanced signal processing techniques can significantly improve the performance of SPC-based systems, as demonstrated in Section \ref{PerfEnhance}. However, the majority of SPC-based OWC systems reported in the literature still rely on conventional signal processing techniques, which fall short on achieving optimal performance. Techniques, such as VNT and pre/post-distortion, that adapt to the unique characteristics of SPC receivers are crucial for unlocking the true potential of this technology. Unfortunately, the lack of accurate photon-counting models and the highly complex nonlinear behaviour of the SPC receivers render the development of truly optimal signal processing techniques especially challenging. 
    To this end, data-driven approaches based on machine learning would be promising for optimizing the link performance of such systems.}

	\subsection{Modulation Schemes}
	Many studies on SPC-based OWC systems in the literature have employed OOK as the standard modulation scheme, mainly because of the limited receiver dynamic range. Recently, the development of large SPAD arrays with low dead time has greatly extended the dynamic range of SPAD-based receivers. This, in turn, has motivated researchers to explore the use of higher-order modulation schemes in SPAD OWC, such as optical OFDM \cite{Huang:22}.  
	Recent research has demonstrated that return-to-zero OOK with appropriate duty cycles outperforms the traditional non-return-to-zero  OOK for SPAD OWC \cite{Zimmermann17}, illustrating the potential for performance enhancement through new modulation designs. Further investigation is necessary to explore the design of spectrally efficient modulation schemes leveraging the unique nonlinear properties of SPC detectors.

\bibliographystyle{IEEEtran}
\bibliography{IEEEabrv,manuscript_final}

\begin{IEEEbiographynophoto}{Shenjie Huang} received the Ph.D. degree in electrical engineering from The University of Edinburgh, U.K., in 2018. He is currently a Research Associate with the Institute for Digital Communications, The University of Edinburgh. 
\end{IEEEbiographynophoto}

\begin{IEEEbiographynophoto}{Danial Chitnis} joined the School of Engineering at The University of Edinburgh as a Chancellor's Fellow in 2017 and is now a Lecturer in Electronics developing detector arrays and systems for a variety of applications from quantum physics to consumer cameras.
\end{IEEEbiographynophoto}

\begin{IEEEbiographynophoto}{Cheng Chen}	
received the Ph.D. degree in electrical engineering from the University of Edinburgh, Edinburgh, U.K., in 2017. He is currently employed as a Research Associate with the Li-Fi Research and Development Centre, The University of Strathclyde. 
\end{IEEEbiographynophoto}

\begin{IEEEbiographynophoto}{Harald Haas}
is the Director of the LiFi Research and Development Centre at the University of Strathclyde. He is co-founder and Chief Scientific Officer of pureLiFi Ltd. 
He is a Humboldt Research Award recipient, a Fellow of the Royal Academy of Engineering, the Royal Society of Edinburgh, the IEEE and the IET.
\end{IEEEbiographynophoto}

\begin{IEEEbiographynophoto}{Mohammad-Ali Khalighi}	
is an Associate Professor with École Centrale Méditérranée, and head of ``Optical Communications for IoT'' group at the Fresnel Institute. 
He serves as the chair of the COST Action CA19111 NEWFOCUS, and as Editor-at-Large for the IEEE Transactions on Communications. 
\end{IEEEbiographynophoto}

\begin{IEEEbiographynophoto}{Robert K. Henderson}	is a Professor of Electronic Imaging in the School of Engineering at the University of Edinburgh. 
In 2014, he was awarded a prestigious ERC advanced fellowship. He is an advisor to Sense Photonics and a Fellow of the IEEE and the Royal Society of Edinburgh.
\end{IEEEbiographynophoto}

\begin{IEEEbiographynophoto}{Majid Safari} is a Professor of Optical and Wireless Communications at the University of Edinburgh and an associate editor of IEEE Transactions on Communications. His main research interest is the application of information theory and signal processing in optical wireless communications. He is a recipient of Mitacs Fellowship and prestigious grants from Leverhulme Trust and EPSRC.
\end{IEEEbiographynophoto}

\end{document}